\documentclass[aps,preprint,showpacs,preprintnumbers,showkeys,eqsecnum,amsmath,amssymb]{revtex4}

\usepackage{graphics}
\usepackage{graphicx}
\usepackage{txfonts}
\usepackage{dcolumn}
\usepackage{mathrsfs}
\usepackage{bm}
\usepackage{amsmath,amssymb,epsfig,float}

\newcommand{\bq}{\begin{equation}}
\newcommand{\eq}{\end{equation}}
\newcommand{\bqn}{\begin{eqnarray}}
\newcommand{\eqn}{\end{eqnarray}}

\begin{document}

\title{ Geometric optics for a coupling model of electromagnetic and gravitational fields}

\author{ Jiliang {Jing}\footnote{Electronic address:
jljing@hunnu.edu.cn}}
\author{Songbai Chen}
\author{Qiyuan Pan}
\affiliation{Department of Physics, Key Laboratory of Low Dimensional Quantum Structures and Quantum Control of Ministry of Education, and Synergetic Innovation Center for Quantum Effects and Applications, Hunan Normal University, Changsha, Hunan 410081, P. R. China.
}

\begin{abstract}

The coupling between the electromagnetic and gravitational fields  results in ``faster than light" photons and  invalids the Lorentz invariance and some laws of physics. A typical example is that the first and third laws of geometric optics are invalid in the usual spacetime.
By introducing an effective spacetime, we find that the wave vector can be casted into null and then it obeys the geodesic equation, the polarization vector is perpendicular to the rays, and the number of photons is conserved. That is to say, the laws of geometric optics are still valid for the modified theory
in the effective spacetime. We also show that the focusing theorem of light rays for the modified theory in the effective spacetime takes the same form as usual.

\end{abstract}

\pacs{03.50.De, 04.60.+v, 41.20.Jb.}

\keywords{modified theory, geometric optics, focusing theorem.}

\maketitle

\section{Introduction}

It is well known that the propagation of light and radio waves are
subject to the laws of geometric optics. The fundamental laws of
geometric optics are: (1) light rays are null geodesics; (2) the
polarization vector is perpendicular to the rays and
parallel-propagated along the rays; and (3) the amplitude is
governed by an adiabatic invariant which, in the quantum language,
states that the number of photons is conserved. The conditions under
which these laws hold are defined by conditions on three lengths:
(1) the typical reduced wavelength of the waves, $\lambda$, as
measured in a typical local Lorentz frame; (2) the typical length
$\mathfrak{L}$ over which the amplitude, polarization, and
wavelength of the waves vary, e.g., the radius of curvature of a
wavefront; (3) the typical radius of curvature $\mathfrak{R}$ of the
spacetime through which the waves propagate. Geometric optics is
valid whenever the reduced wavelength is very short compared to each
of the other scales. We should note that these laws, regardless of
in the flat or curved spacetime, are derived from the usual free
Maxwell field.

Recent investigations show that the interaction between the
electromagnetic and gravitational fields could be appeared naturally
in quantum electrodynamics with the photon effective action
originating from one-loop vacuum polarization in curved spacetime
\cite{Drummond}.  This coupling model of the electromagnetic and
gravitational fields is of great interest, since the appearance of
cross-terms in the Lagrangian leads to the modifications of the
coefficients involving the higher-order derivatives both in the
Maxwell and Einstein equations. So the electromagnetic theories
containing this coupling term have been studied extensively
\cite{Drummond,Balakin,Faraoni,Hehl}. Drummond and Hathrell
\cite{Drummond} argued that the quantum corrections introduce tidal
gravitational forces on the photons which alter the characteristics
of propagation, so that in some cases photons travel at speeds
greater than unity. The one-loop effective action for QED in curved
spacetime contains the equivalence principle violating interactions
between the electromagnetic field and the spacetime curvature, and
these interactions lead to a dependence of the photon velocity on
the motion and polarization
directions\cite{Drummond,Dan,Cai,Shore,Cho,Dalvit}. By taking the
analogy between the eikonal equation in geometric optics and the
particle equation of motion, Ahmadi and Nouri-Zonoz \cite{Ahmadi}
investigated the phase structure and the trajectory of the
propagating photon semiclassically in the modified theory.

Although the coupling models for the electromagnetic and gravitational fields have been extensive used recently,  they will result in ``faster than light" photons and  will invalid the Lorentz invariance and some physical laws. For this coupling models, the characteristics of propagation of the light are altered and the
laws of geometric optics, in general, are invalid in the usual
spacetime. The reason is that the electromagnetic fields are
described by the modified theories rather than the usual free
Maxwell theory now. In this paper, we want to show that, by
introducing an effective spacetime, the laws of geometric optics and
focusing theorem of light rays are still valid for the modified
theory.

The plan of the paper is as follows. In the next section we
introduce a model for the interaction between the electromagnetic
and gravitational fields, and obtain the motion equations of light
by using the geometric optics approximation in the usual spacetime.
In section III, by using  the effective spacetime introduced in the
appendix, we investigate the laws of geometric optics and the
focusing theorem of light rays. We present our conclusions in the
last section.

\section{Modified theory for couplings between electromagnetic and gravitational fields}

To study the interactions between the electromagnetic and
gravitational fields, it is natural to consider the couplings
between the Maxwell field and the Weyl tensor. Therefore, the
electromagnetic theory with the Weyl correction has been studied
extensively \cite{Weyl1,Wu2011,Ma2011,Momeni,Roychowdhury,zhao2013}.
The action for a toy model of the electromagnetic field coupled to
the Weyl tensor can be expressed as  \cite{Weyl1}
\begin{eqnarray}
S=\int d^4 x \sqrt{-g}\bigg[\frac{R}{16\pi
G}-\frac{1}{4}\bigg(F_{\mu\nu}F^{\mu\nu}-4\alpha
C^{\mu\nu\rho\sigma}F_{\mu\nu}F_{\rho\sigma}\bigg)\bigg], \label{acts}
\end{eqnarray}
where $F_{\mu\nu}=A_{\nu;\mu}-A_{\mu;\nu}$ is the usual
electromagnetic tensor with a vector potential $A_\mu$, $\alpha$ is
the coupling constant which has the dimension of the length-squared,
and $C_{\mu\nu\rho\sigma}$ is the Weyl tensor defined by
\begin{eqnarray}
C_{\mu\nu\rho\sigma}=R_{\mu\nu\rho\sigma}-(
g_{\mu[\rho}R_{\sigma]\nu}-g_{\nu[\rho}R_{\sigma]\mu})+\frac{1}{3}R
g_{\mu[\rho}g_{\sigma]\nu},
\end{eqnarray}
where $g_{\mu\nu}$ represents the usual spacetime and the brackets
around indices refer to the antisymmetric part. Varying the action
(\ref{acts}) with respect to $A_{\mu}$, we obtain
\begin{eqnarray}
\nabla_\mu(F^{\mu\nu}-4\alpha C^{\mu\nu\rho\sigma}F_{\rho
\sigma})=0. \label{main 1}
\end{eqnarray}
Under the geometric optics assumption, we can set
\begin{eqnarray}
A^{\mu}=(a^\mu+\varepsilon b^\mu +\varepsilon^2 c^\mu+......)e^{i\theta}, \label{main 2}
\end{eqnarray}
where $\theta$ is a phase, and $\varepsilon=\lambda/L$ is a small
dimensionless number with the minimum $L$ of $\mathfrak{L}$ and
$\mathfrak{R}$. Substituting Eq. (\ref{main 2}) into Eq. (\ref{main
1}) and noting that the polarization vector is perpendicular to the
rays, we get from the leading term (order $1/\varepsilon^2$)
\begin{eqnarray}
k_\mu k^{\mu} a^\nu+8 \alpha C^{\mu\nu\rho\sigma}k_\mu k_\sigma a_\rho=0. \label{main 3}
\end{eqnarray}
If we take
\begin{eqnarray}
a^\mu=a f^\mu, \label{main 4}
\end{eqnarray}
where $f^\mu$ is a unit vector along $a^\mu$, Eq. (\ref{main 3}) can be rewritten as
\begin{eqnarray}
k_\mu (k^{\mu} +8 \alpha C^{\mu\nu\rho\sigma} k_\sigma f_\rho
f_\nu)=0, \label{main 5}
\end{eqnarray}
which tells us that the wave vector is not null now. That is to say,
it is different from the null wave vector $k_\mu k^\mu=0$ in the
usual free Maxwell theory.

From the subleading term (order $1/\varepsilon$), we can obtain the
propagation equation for the vector amplitude
\begin{eqnarray}
2 a^\nu_{;\mu}k^\mu+(k^\mu_{\ ;\mu} a^\nu-a^\mu_{\ ;\mu}k^\nu)= 8
\alpha [(C^{\mu\nu\rho\sigma} a_\sigma
k_\rho)_{;\mu}+C^{\mu\nu\rho\sigma} a_{\sigma;\rho} k_\mu],
\label{main 6a}
\end{eqnarray}
which can be reformulated as a conservation law
\begin{eqnarray}
[a^2 (k^{\mu} +8 \alpha C^{\mu\nu\rho\sigma} k_\sigma f_\rho f_\nu)]_{;\mu}=0. \label{main 6}
\end{eqnarray}
Consequently, the vector  $a^2 (k^{\mu} +8 \alpha
C^{\mu\nu\rho\sigma} k_\sigma f_\rho f_\nu)$ can be considered as a
conserved current, and the integral
\begin{eqnarray}
\oint[a^2 (k^{\mu} +8 \alpha C^{\mu\nu\rho\sigma} k_\sigma f_\rho f_\nu)]d^3\Sigma_\mu, \label{main 7}
\end{eqnarray}
has a fixed and unchanging value for each 3-volume cutting a given tube formed of light rays.

\section{geometric optics for modified theory in effective spacetime}

Eq. (\ref{main 5}) shows that the wave vector is not null, which
means that the first law of geometric optics is invalid for the
modified theory in the usual spacetime. And Eq. (\ref{main 6}) or
(\ref{main 7}) tells us that the conserved current $a^2 (k^{\mu} +8
\alpha C^{\mu\nu\rho\sigma} k_\sigma f_\rho f_\nu)$ is directly related to
the curvature of the usual spacetime in the modified theory.  In the
following we will prove that all laws of geometric optics are valid
for the modified theory in the effective spacetime.

\subsection{Light rays are null geodesics in the effective spacetime}

In the usual free Maxwell theory, we do not need to distinguish
between the photon 4-velocity $U^\mu$, i.e., the tangent vector to
the light rays, and the wave vector $k_\mu$ since they are simply
related by using the spacetime metric $g^{\mu\nu}$ to raise the
index, i.e., $U^\mu\sim k^\mu=g^{\mu\nu}k_\nu$ (the detail proof
please see the footnote \footnotemark{ }\footnotetext{From $k_\mu k^\mu=0$ we can easily get
$k_{\nu;\mu}k^\mu=0$. On the other hand, the wave vector $k_\nu$
satisfies the geodesic equation $\frac{D k_\nu}{d \lambda}=\frac{d
k_\nu}{d\lambda}-\Gamma^\sigma_{\nu\mu}k_\sigma U^\mu =
k_{\nu;\mu}U^\mu=0$. Therefore, we have $k^\mu \sim U^\mu$.}).

However, there is an important distinction in the modified theory.
The wave vector $k_\mu$, defined as the derivative of the phase, is
a covariant vector, whereas the 4-velocity $U^\mu$ is a true
contravariant vector. The relation between them is nontrivial.

In order to cast the results obtained in the modified theory into
the familiar form, we introduce the new effective metric
$\mathfrak{g}_{\mu\nu}$ (please see appendix A) and define
\begin{eqnarray}
\tilde{k}^\mu=\mathfrak{g}^{\mu\nu}k_\nu, \ \ \ \ \ \ \  \  \tilde{k}_\mu =k_\mu. \label{main 8}
\end{eqnarray}
Then, we can write the light cone condition (\ref{main 5}) for the
wave vector as the homogeneous form
\begin{eqnarray}
\tilde{k}^\mu \tilde{k}_\mu=\mathfrak{g}^{\mu\nu}\tilde{k}_\mu \tilde{k}_\nu=\mathfrak{g}_{\mu\nu}\tilde{k}^\mu \tilde{k}^\nu=0. \label{main 9}
\end{eqnarray}

In the effective spacetime, we can prove that $\tilde{k}^\mu \sim
\tilde{U}^\mu$ with $\tilde{U}^\mu=\frac{d\tilde{x}^\mu}{d\lambda}$.
Using the fact that $\tilde{k}_\mu=\theta_{,\mu}$ is the gradient of
a scalar and $\theta_{;\mu\nu}=\theta_{;\nu\mu}$, from
$\tilde{k}^\mu \tilde{k}_\mu =0$ we can get the propagation equation
for the wave vector
\begin{eqnarray}
\tilde{k}_{\nu;\mu}\tilde{k}^\mu=0. \label{main 10}
\end{eqnarray}
On the other hand, the wave vector $\tilde{k}_\nu$ satisfies the
geodesic equation
\begin{eqnarray}
\frac{D \tilde{k}_\nu}{d \lambda}=\frac{d \tilde{k}_\nu}{d\lambda}-\mathfrak{\Gamma}^\sigma_{\nu\mu}\tilde{k}_\sigma \tilde{U}^\mu=\tilde{k}_{\nu;\mu}\tilde{U}^\mu=0.\label{main 11}
\end{eqnarray}
By comparing Eqs. (\ref{main 10}) and (\ref{main 11}), we find that $\tilde{k}^\mu \sim \tilde{U}^\mu$.

From the geodesic equation (\ref{main 10}) (or (\ref{main 11})) and
the light cone condition (\ref{main 5}) which are derived from the
modified theory in the effective spacetime, we can state that the
light rays are null geodesics now. That is to say, the first law of
geometric optics is still valid for the modified theory in the
effective spacetime.

\subsection{Law of conservation of photon number in the effective spacetime}

By using Eq. (\ref{main 8}), in the effective spacetime we can
reformulate (\ref{main 6}) as
\begin{eqnarray}
(a^2 \tilde{k}^{\mu})_{;\mu}=0. \label{main 6b}
\end{eqnarray}
The vector  $a^2 \tilde{k}^{\mu} $ is the conserved current now and the integral
\begin{eqnarray}
\oint(a^2 \tilde{k}^{\mu})d^3\Sigma_\mu, \label{main 7a}
\end{eqnarray}
has a fixed and unchanging value for each 3-volume cutting a given
tube formed of light rays. In the integral the tube must be so
formed of rays that an integral of $a^2\tilde{k}^\mu$ over the walls
of the tube will give zero. What is the physical significance of Eq.
(\ref{main 7a})? To remain a purely classical one, it tells us that
the ``number of light rays" is conserved and $a^2\tilde{k}^0$ is the
``density of light rays" on an $x^0=constant$ hypersurface. However,
for the concrete physical interpretation, we prefer to consider Eq.
(\ref{main 7a}) as the law of conservation of photon number along
this tube.

\subsection{Focusing theorem in the effective spacetime}

For a bundle of rays lying in a surface of constant phase in the
tube, we can take a tiny area of the  two-dimensional cross section
as $\sigma$. Then the conserved equation (\ref{main 7a}) can be
reformulated as
\begin{eqnarray}
a^2 \sigma=constant, \label{main 12}
\end{eqnarray}
i.e.,
\begin{eqnarray}
\frac{d(a^2 \sigma)}{d\lambda}=(a^2 \sigma)_{;\mu}\tilde{k}^{\mu}=0. \label{main 12a}
\end{eqnarray}
Using Eqs. (\ref{main 6b}) and (\ref{main 12a}), we can show that
the area changes from point to point along the bundle of rays as a
result of the rays which diverge away from each other or converge
toward each other
\begin{eqnarray}
\sigma_{;\mu}\tilde{k}^{\mu}=\sigma \tilde{k}^{\mu}_{;\mu}. \label{main 13}
\end{eqnarray}
Then, we can find that
\begin{eqnarray}
\frac{d^2\sqrt{\sigma}}{d\lambda^2}=-\left(\delta^2+\frac{1}{2}
\mathfrak{R}_{\mu\nu}\tilde{k}^{\mu}\tilde{k}^{\nu}
\right)\sqrt{\sigma},  \label{main 14}
\end{eqnarray}
with
\begin{eqnarray}
 \delta^2=\frac{1}{2}\tilde{k}^{\mu;\nu}\tilde{k}_{\nu;\mu}
 -\frac{1}{4}(\tilde{k}^{\mu}_{;\mu})^2, \label{main 14a}
\end{eqnarray}
where $\mathfrak{R}_{\mu\nu}$ is the Ricci curvature tensor of the effective
spacetime, and the quantity $\delta$ is the shear of the bundle of
rays because it measures the extent to which neighboring rays are
sliding past each other. Therefore, the focusing equation (\ref{main
14}) tells us that the shear focuses on a bundle of rays, and the
spacetime curvature also focuses on it if $\mathfrak{R}_{\mu\nu}
\tilde{k}^{\mu} \tilde{k}^{\nu}>0$, but defocuses on it if
$\mathfrak{R}_{\mu\nu}\tilde{k}^{\mu} \tilde{k}^{\nu}<0$.

Assuming that the energy density is non-negative in the effective
spacetime, from the focusing equation (\ref{main 14}) and the
Einstein field equations we obtain the focusing theorem
\begin{eqnarray}
\frac{d^2\sqrt{\sigma}}{d\lambda^2}\leq 0, \label{main 15}
\end{eqnarray}
which takes the same form as the usual free Maxwell theory.

\section{conclusions}

In the coupling models for the electromagnetic and gravitational fields, which  are extensive used recently, the wave vector is not null and the
conserved current $a^2 (k^{\mu} +8 \alpha C^{\mu\nu\rho\sigma}
k_\sigma  f_\rho f_\nu)$ is directly related to the curvature of the usual spacetime. That is to say, the first and third laws of geometric optics are invalid in the modified
theory due to the ``faster than light" photons and  invalid of the Lorentz invariance.

By introducing the effective spacetime, we first show that the wave
vector becomes null and obeys the geodesic equation, which means that the
first law of geometric optics is valid for the modified theory
in the effective spacetime. Noting that the integral $ \oint(a^2
\tilde{k}^{\mu})d^3\Sigma_\mu$ has a fixed and unchanging value for
each 3-volume cutting a given tube formed of light rays, we then
find that the amplitude is governed by an adiabatic invariant which,
in the quantum language, states that the number of photons is
conserved.

From the focusing equation we know that the shear focuses on a
bundle of rays, and the spacetime curvature also focuses on it if
$\mathfrak{R}_{\mu\nu} \tilde{k}^{\mu} \tilde{k}^{\nu}>0$, but
defocuses on it if $\mathfrak{R}_{\mu\nu}\tilde{k}^{\mu}
\tilde{k}^{\nu}<0$. Furthermore, by using the focusing equation and
the Einstein field equations, we find that, if the energy density is
non-negative, the focusing theorem of light rays for the modified
theory in the effective spacetime takes the same form as usual.

\appendix

\section{ Effective metrics }

For simplicity and clarity, we just consider a general
four-dimensional static and spherically symmetric spacetime. The
metric can be expressed as
\begin{eqnarray}
ds^2&=&g_{00} dt^2+g_{11} dr^2+g_{22}(
d\theta^2+sin^2\theta d\phi^2),\label{m1}
\end{eqnarray}
where $g_{00}$, $g_{11}$ and $g_{22}$ are functions of the polar
coordinate $r$ only. For metric (\ref{m1}) the appropriate basis
1-forms are
\begin{eqnarray}
e^0=e^0_t d t,~~~e^1=e^1_r d r,~~~e^2=e^2_\theta d
\theta,~~~e^3=e^3_\phi d \phi, \label{e1}
\end{eqnarray}
where the vierbein fields $e^a_\mu$ defined by
\begin{eqnarray}
g_{\mu\nu}=\eta_{ab}e^a_{\mu}e^b_{\nu},
\end{eqnarray}
here $\eta_{ab}$ is the Minkowski metric with the signature $(-,+,+,+)$.  The vierbeins for metric (\ref{m1}) are
\begin{eqnarray}
e^a_{\mu}=diag(\sqrt{-g_{00} },\;\sqrt{g_{11} },\;\sqrt{g_{22} } ,\;\sqrt{g_{33} }).
\end{eqnarray}

Using the antisymmetric combination of vierbeins \cite{Drummond}
\begin{eqnarray}
U^{ab}_{\mu\nu}=e^a_{\mu}e^b_{\nu}-e^a_{\nu}e^b_{\mu},
\end{eqnarray}
the Weyl tensor can be rewritten as
\begin{eqnarray}
C_{\mu\nu\rho\sigma}=\mathcal{A}\bigg(2U^{01}_{\mu\nu}U^{01}_{\rho\sigma}-
U^{02}_{\mu\nu}U^{02}_{\rho\sigma}-U^{03}_{\mu\nu}U^{03}_{\rho\sigma}
+U^{12}_{\mu\nu}U^{12}_{\rho\sigma}+U^{13}_{\mu\nu}U^{13}_{\rho\sigma}-
2U^{23}_{\mu\nu}U^{23}_{\rho\sigma}\bigg),
\end{eqnarray}
with
\begin{eqnarray}
\mathcal{A}&=&-\frac{1}{12 (g_{00} g_{11} )^2 g_{22}
}\bigg\{\left[g_{00} g_{11} g_{00} ''-\frac{1}{2}(g_{00} g_{11}
)'g_{00} '\right]g_{22}-g_{00} ^2g_{11} g_{22} '' \nonumber \\
&&+\frac{1}{2}(g_{00} ^2g_{11} '-g_{00} g_{11} g_{00}
')g_{22}'-2(g_{00} g_{11} )^2\bigg\}.
\end{eqnarray}
In order to cast the equation of motion for the coupled photon
propagation into a simplified form, we introduce three linear
combinations of momentum components \cite{Drummond}
\begin{eqnarray}
l_{\nu}=k^{\mu}U^{01}_{\mu\nu},~~~~~~~~
n_{\nu}=k^{\mu}U^{02}_{\mu\nu},~~~~~~~~
m_{\nu}=k^{\mu}U^{23}_{\mu\nu},
\end{eqnarray}
together with the dependent combinations
\begin{eqnarray}
&&p_{\nu}=k^{\mu}U^{12}_{\mu\nu}=\frac{1}{e^0_0 k^0}\bigg(e^1_1 k^1 n_{\nu}-e^2_2\tilde{k}^2 l_{\nu}\bigg),\nonumber\\
&&r_{\nu}=k^{\mu}U^{03}_{\mu\nu}=\frac{1}{e^2_2 \tilde{k}^2}\bigg(e^0_0 k^0 m_{\nu}+e^3_3k^3 l_{\nu}\bigg),\nonumber\\
&&q_{\nu}=k^{\mu}U^{13}_{\mu\nu}=\frac{e^1_1 k^1}{e^2_2 \tilde{k}^2}m_{\nu}+
\frac{e^1_1 e^3_3 k^1k^3}{e^0_0 e^2_2 k^0 \tilde{k}^2}n_{\nu}-\frac{e^3_3 k^3}{e^0_0 k^0}l_{\nu}.\label{vect3}
\end{eqnarray}
The vectors $l_{\nu}$, $n_{\nu}$, $m_{\nu}$ are independent and
orthogonal to the wave vector $k_{\nu}$. Contracting Eq. (\ref{main
3}) with $l_{\nu}$, $n_{\nu}$, $m_{\nu}$ respectively, using the
relation (\ref{vect3}) and introducing three independent
polarisation components $(a\cdot l)$, $(a\cdot n)$, and $(a\cdot
m)$, we find that the equation of motion of the photon coupling with
the Weyl tensor can be simplified as
\begin{eqnarray}
\left(\begin{array}{ccc}
K_{11}&0&0\\
K_{21}&K_{22}&
K_{23}\\
0&0&K_{33}
\end{array}\right)
\left(\begin{array}{c}
a \cdot l\\
a \cdot n
\\
a \cdot m
\end{array}\right)=0,\label{Kk}
\end{eqnarray}
with the coefficients
\begin{eqnarray}
K_{11}&=&(1+16\alpha \mathcal{A})(g^{00}k_0k_0+g^{11}k_1k_1)+(1-8\alpha \mathcal{A})(g^{22}k_2k_2+g^{33}k_3k_3),\nonumber\\
K_{22}&=&(1-8\alpha \mathcal{A})(g^{00}k_0k_0+g^{11}k_1k_1+g^{22}k_2k_2+g^{33}k_3k_3),
\nonumber\\
K_{21}&=&24\alpha \mathcal{A} \sqrt{g^{11}g^{22}}k_1k_2,\;\;\;\;\;\;\;\;\;\;
K_{23}=8\alpha \mathcal{A}\sqrt{-g^{00}g^{33}}k_0k_3,\nonumber\\
K_{33}&=&(1-8\alpha \mathcal{A})(g^{00}k_0k_0+g^{11}k_1k_1)+(1+16\alpha \mathcal{A})(g^{22}k_2k_2+g^{33}k_3k_3). \nonumber \\
\end{eqnarray}
The condition of Eq. (\ref{Kk}) with the non-zero solution is
$K_{11}K_{22}K_{33}=0$. The first root $K_{11}=0$ leads to the
modified light cone
\begin{eqnarray}
(1+16\alpha \mathcal{A})(g^{00}k_0k_0+g^{11}k_1k_1)+(1-8\alpha \mathcal{A})(g^{22}k_2k_2+g^{33}k_3k_3)=0, \label{Kk31}
\end{eqnarray}
which corresponds to the case where the polarisation vector
$a_{\mu}$ is proportional to $l_{\mu}$. The second root $K_{22}=0$
corresponds to an unphysical polarisation and should be neglected.
The third root is $K_{33}=0$, i.e.,
\begin{eqnarray}
(1-8\alpha \mathcal{A})(g^{00}k_0k_0+g^{11}k_1k_1)+(1+16\alpha \mathcal{A})(g^{22}k_2k_2+g^{33}k_3k_3)=0,\label{Kk32}
\end{eqnarray}
which means that the vector $a_{\mu}=\lambda m_{\mu}$.

The above discussions show that the light cone condition depends on
not only the coupling between the photon and the Weyl tensor, but
also on the polarizations. We know from results (\ref{Kk31}) and
(\ref{Kk32}) that the light cone condition is not modified for the
radially directed photons (i.e., $k_2=k_3=0$) but is modified for
the orbital photons (i.e., $k_1=k_2=0$), and the velocities of the
photons for the two polarizations are different, i.e., the
phenomenon of gravitational birefringence \cite{Drummond}.

In order to cast the light cone condition (\ref{main 5}) ( or the result (\ref{Kk31}))  obtained in the modified theory into the familiar form, we introduce a new effective contravariant metric $\mathfrak{g}^{\mu\nu}$ in which the nonzero  components are
\begin{eqnarray}
&&\mathfrak{g}^{00}=(1+16\alpha \mathcal{A})g^{00}, \nonumber \\ &&\mathfrak{g}^{11}=(1+16\alpha \mathcal{A})g^{11}, \nonumber\\
&&\mathfrak{g}^{22}=(1-8\alpha \mathcal{A})g^{22}, \nonumber \\ &&\mathfrak{g}^{33}=(1-8\alpha \mathcal{A})g^{33}, \label{eff m1}
\end{eqnarray}
and its covariant metric is given by
\begin{eqnarray}
&&\mathfrak{g}_{00}=\frac{1}{1+16\alpha \mathcal{A}}g_{00},  \nonumber \\ && \mathfrak{g}_{11}=\frac{1}{1+16\alpha \mathcal{A}}g_{11}, \nonumber\\
&&\mathfrak{g}_{22}=\frac{1}{1-8\alpha \mathcal{A}}g_{22}, \nonumber
\\ && \mathfrak{g}_{33}=\frac{1}{1-8\alpha \mathcal{A}}g_{33}.
\label{eff m1a}
\end{eqnarray}
Then, defining
\begin{eqnarray}
\tilde{k}^\mu=\mathfrak{g}^{\mu\nu}k_\nu, \ \ \ \ \ \ \  \
\tilde{k}_\mu =k_\mu, \label{main 8a}
\end{eqnarray}
we can write the light cone condition (\ref{main 5}) for the result
(\ref{Kk31}) into the homogeneous form
\begin{eqnarray}
\tilde{k}^\mu \tilde{k}_\mu=\mathfrak{g}^{\mu\nu}\tilde{k}_\mu \tilde{k}_\nu=\mathfrak{g}_{\mu\nu}\tilde{k}^\mu \tilde{k}^\nu=0. \label{main 9a}
\end{eqnarray}

Similarly, in order to cast the light cone condition (\ref{main 5})
for the result (\ref{Kk32}) into the homogeneous form, we should
introduce the effective contravariant metric
\begin{eqnarray}
&&\mathfrak{g}^{00}=(1-8\alpha \mathcal{A})g^{00},  \nonumber \\
&&\mathfrak{g}^{11}=(1-8\alpha \mathcal{A})g^{11}, \nonumber\\
&&\mathfrak{g}^{22}=(1+16\alpha \mathcal{A})g^{22},  \nonumber \\ &&
\mathfrak{g}^{33}=(1+16\alpha \mathcal{A})g^{33}, \label{eff m2}
\end{eqnarray}
and its covariant metric
\begin{eqnarray}
&&\mathfrak{g}_{00}=\frac{1}{1-8\alpha \mathcal{A}}g_{00},  \nonumber \\ && \mathfrak{g}_{11}=\frac{1}{1-8\alpha \mathcal{A}}g_{11}, \nonumber\\ &&\mathfrak{g}_{22}=\frac{1}{1+16\alpha \mathcal{A}}g_{22},  \nonumber \\ && \mathfrak{g}_{33}=\frac{1}{1+16\alpha \mathcal{A}}g_{33}.
\label{eff m2a}
\end{eqnarray}
Similarly, we can also cast the light cone condition (\ref{main 5}) for the result
(\ref{Kk32}) into the homogeneous form.

\begin{acknowledgments}

This work is supported by the  National Natural Science Foundation
of China under Grant Nos. 11475061;  the SRFDP under Grant No.
20114306110003;  S. Chen's work was
partially supported by the National Natural Science Foundation of
China under Grant No. 11275065;  and Q. Pan's work was partially
supported by the National Natural Science Foundation of China under
Grant No. 11275066.

\end{acknowledgments}


\end{document}